# On the possibility of resonant acceleration of charged particles by the field of a transverse electromagnetic wave in vacuum

V. A. Buts

National Scientific Center "Kharkov Institute of Physics and Technology," National Academy of Sciences of Ukraine, 61108 Kharkov, Ukraine and Radio Astronomy Institute, National Academy of Sciences of Ukraine, 61002, Kharkov, Ukraine

Electronic mail: vbuts@kipt.kharkov.ua

The analytical and numerical analysis of the dynamics of charged particles in the field of an intensive transverse electromagnetic wave in vacuum presented in the article. Identifies the conditions for resonant acceleration of particles. These conditions are formulated. The features and the mechanism of this acceleration are discussed.

1. **Introduction**

The acceleration of charged particles by intense electromagnetic fields in a vacuum is a tempting goal of many studies. The advantages of this acceleration are well known. The difficulties encountered by researchers in trying to achieve this goal are also known. It is sufficient to mention the publications [1-3] and the literature cited there to understand the challenges, importance of the problem and the current state of the research in this area. Here we present the results of the analytical and numerical analysis of the dynamics of a charged particle in the field of intensive electromagnetic wave in vacuum and identify the conditions of resonant interaction of particles and the field. The work consists of four sections and a conclusion. The second section describes the dynamics of particles in the field of a circularly polarized wave showing the possibility of resonant acceleration of the particles by the wave. The third section describes the results of the numerical analysis of this dynamics, illustrating the resonant acceleration numerically. The fourth section deals with the dynamics of particles in a field with linear polarization. In the conclusion, the main results are formulated and some of them are discussed.

2. **The problem and main equations**

Consider the problem of the dynamics of a charged particle that moves in the field of a plane electromagnetic wave, which generally has the following components:

$$\mathcal{E} = \mathrm{Re}(\mathbf{E}\exp(i\omega t - i\mathbf{kr})),$$
$$\mathbf{H} = \mathrm{Re}\left(\frac{c}{\omega}[\mathbf{kE}]\exp(i\omega t - i\mathbf{kr})\right) \quad (1)$$

Where $\mathbf{E} = \boldsymbol{\alpha}\cdot E$, $\boldsymbol{\alpha} = \{\alpha_x, i\alpha_y, \alpha_z\}$ is the wave polarization vector.

Without loss of generality, we can assume that the wave vector k has only two nonzero components $k_x$ and $k_z$. In dimensionless variables $\mathbf{p} \to \mathbf{p}/mc$, $\tau \to \omega t$, $\mathbf{r} \to \mathbf{r}(\omega/c)$ the equations of motion of the particle can be transformed to the form.

$$\frac{d\mathbf{p}}{d\tau} = \left(1 - \frac{\mathbf{kp}}{\gamma}\right)\mathrm{Re}\left(\mathcal{E}e^{i\psi}\right) + \frac{\mathbf{k}}{\gamma}\mathrm{Re}\left[(\mathbf{p}\mathcal{E})e^{i\psi}\right] \quad (2)$$

$$\mathbf{v} = \frac{d\mathbf{r}}{d\tau} = \frac{\mathbf{p}}{\gamma}, \quad \dot{\psi} = \frac{d\psi}{d\tau} = 1 - \frac{\mathbf{kp}}{\gamma},$$

Where $\mathcal{E} = e\mathbf{E}/mc\omega$, $\psi = \tau - \mathbf{kr}$, $\mathbf{k}$ - unit vector in the direction of wave propagation, $\gamma = \sqrt{1 + \mathbf{p}^2}$ is the particle energy, $\mathbf{p}$ - is its momentum.

Multiplying the first of equation of (2) by $\mathbf{p}$, we obtain the following equation describing the change in the particle energy:

$$\frac{d\gamma}{d\tau} = \mathrm{Re}\left(\mathbf{v}\mathcal{E}e^{i\psi}\right) \quad (3)$$

Using equations (3), from the system of equations (2) we find the integral of motion:

$$\mathbf{p} - \text{Re}(i\boldsymbol{\mathcal{E}} e^{i\psi}) - \mathbf{k}\gamma = \mathbf{const} \equiv \mathbf{C} \quad (4)$$

## 2. Dynamics of particles in the field of a circularly polarized wave

The dynamics of particles in the field of a circularly polarized wave most fully reflects all the features of the dynamics of interest. Therefore, we consider this dynamic first. We assume that the wave vector of the electromagnetic wave is located in the plane $\{x, z\}$, that is $\mathbf{k} = \{k_x, 0, k_z\}$. Then the system of equations (2) for describing the motion of particles in the field of a wave with circular polarization takes the form:

$$\dot{p}_x = \varepsilon_x(1 - k_x v_x - k_z v_z)\cos\psi +$$
$$+ \frac{k_x}{\gamma}\left[(\varepsilon_x p_x + \varepsilon_z p_z)\cos\psi - \varepsilon_y p_y \sin\psi\right]$$
$$\dot{p}_y = -\varepsilon_y(1 - k_x v_x - k_z v_z)\sin\psi \quad (5)$$
$$\dot{p}_z = \varepsilon_z(1 - k_x v_x - k_z v_z)\cos\psi +$$
$$+ \frac{k_z}{\gamma}\left[(\varepsilon_x p_x + \varepsilon_z p_z)\cos\psi - \varepsilon_y p_y \sin\psi\right]$$
$$\dot{\gamma} = \frac{1}{\gamma}\left[(\varepsilon_x p_x + \varepsilon_z p_z)\cos\psi - \varepsilon_y p_y \sin\psi\right]$$

Here $\dot{p} \equiv \dfrac{dp}{d\tau}$.

In reality, the system of equations (5) is a system of equations that describes the dynamics of particles in the field of an elliptically polarized wave. However, it is enough for us to consider the dynamics of particles in the field of a circularly polarized wave. Therefore, below we assume that $\varepsilon_z = 0$, $\varepsilon_x = \varepsilon_y = \varepsilon$.

Let us first consider the case when a circularly polarized wave propagates strictly along the axis z ($k_x = \sin\varphi = 0$, $k_z = \cos\varphi = 1$)

$$\dot{p}_x = \dot{\psi} \cdot \varepsilon \cdot \cos\psi$$
$$\dot{p}_y = -\dot{\psi} \cdot \varepsilon \cdot \sin\psi \quad (6)$$

The solution to the system (6) are the functions:

$$p_x = \varepsilon \sin\psi + F_0 = \gamma v_x = \frac{C_z}{\dot{\psi}}\dot{x} = C_z \frac{dx}{d\psi}$$

$$p_y = \varepsilon \cos\psi + \Phi_0 = \gamma v_y = \frac{C_z}{\dot{\psi}}\dot{y} = C_z \frac{dy}{d\psi}$$

$$p_z = p_z(0) + \frac{E^2}{|C_z|}\left[1 - \cos(\psi - \psi_0)\right] \quad (7)$$

Where $F_0 = p_x(0) - \varepsilon \sin\psi_0$,
$\Phi_0 = p_y(0) - \varepsilon \cos\psi_0$, $p_z - \gamma = C_z = const$ - integral.

Zero indices correspond to the initial values of the functions. Note that similar solutions were obtained in [4]. Let's pay attention to expressions (7). It follows from them that the transverse dynamics of particles is proportional to the first power of the wave force parameter, and the longitudinal dynamics is proportional to the square of this parameter.

Thus, the parameter $\varepsilon = 1$ separates qualitatively different particle dynamics. When $\varepsilon < 1$ the transverse (usual) dynamics of particles prevails. If $\varepsilon > 1$, then the longitudinal dynamics of particles prevails. Note also that the second term in the expression for the longitudinal momentum is inversely proportional to the integral $(\gamma - p_z = -C_z = const)$. In many cases, this integral can be small. In this case, the longitudinal motion of particles substantially exceeds the transverse motion.

Consider the structure of the transverse motion of particles. For this, we write down expressions for the transverse coordinates of the particles:

$$(x - x_0) = -\frac{\varepsilon}{C_z}\cos\psi + F_0 \cdot \psi + \frac{1}{C_z}\varepsilon \cdot \cos\psi_0$$

$$(y - y_0) = \frac{\varepsilon}{C_z}\sin\psi + \Phi_0 \cdot \psi - \frac{1}{C_z}\varepsilon \cdot \sin\psi_0 \quad (8)$$

To be specific, we choose the following values of the initial parameters:

$$\psi_0 = 0, \; F_0 = \Phi_0 = 0, \; p_x(0) = 0, \; p_y(0) = \varepsilon \quad (9)$$

Then the trajectories of particles in the field of such a wave become the circles in their cross – section:

$$(x - x_1)^2 + (y - y_0)^2 = \left(\frac{\varepsilon}{C_z}\right)^2 \quad (10)$$

The most interesting here is the fact that the transverse dynamics of particles does not affect the phase dynamics of particles. This effect occurs only when the wave propagates at a certain angle to the z axis, when the transverse wavenumber is nonzero ($k_x = \sin\varphi \neq 0$).

If the transverse component of the wave vector of a wave with circular polarization is nonzero, then such simple expressions that describe the dynamics of particles cannot be obtained. However, if this component of the wave vector is small ($k_x \ll 1$), then it is possible to discover new important features of the particle dynamics. In this case, it is convenient to introduce the new variables:

$$p_x = p_\perp \cos\theta, \quad p_y = p_\perp \sin\theta, \quad p_z = p_\parallel,$$
$$p_\perp = \sqrt{p_x^2 + p_y^2}, \quad x = \xi - \frac{p_\perp}{\gamma\dot\psi}\sin\theta,$$
$$y = \eta + \frac{p_\perp}{\gamma\dot\psi}\cos\theta \quad (11)$$

These new variables explicitly take into account the oscillatory dynamics of particles in the transverse direction. Transverse dynamics and phase dynamics in these new variables are described by the equations:

$$\dot p_\perp = \dot p_x \cos\theta + \dot p_y \sin\theta,$$
$$\dot\theta = \frac{1}{p_\perp}\left(\dot p_y \cos\theta - \dot p_x \sin\theta\right), \quad (12)$$
$$\psi = (\tau - k_z z - k_x \xi) - \mu\sin\theta \equiv a - \mu\sin\theta$$

Where $\mu = \frac{k_x p_\perp}{\gamma \cdot \dot\psi}$.

Below it is convenient to use the expansion formulas (see, for example, [5]):

$$\cos\psi = \cos(a - \mu\sin\theta) = \sum_{n=-\infty}^{\infty} J_n(\mu)\cos(a - n\theta)$$
$$\sin\psi = \sin(a - \mu\sin\theta) = \sum_{n=-\infty}^{\infty} J_n(\mu)\sin(a - n\theta)$$

The condition $k_x \ll 1$; $(k_z - 1) \ll 1$, allows to consider the values $k_x$ present only in the expressions for the phases. Then the second equation (the equation for the phases) of the system (12) takes the form:

$$\dot\theta = -\varepsilon\frac{(1-v_z)}{p_\perp}\sum_{n=-\infty}^{\infty} J_n(\mu)\sin\left[(\tau - z) + \theta - n\theta\right]$$
(13)

Here $\mu \ll 1$, $\gamma\dot\psi \simeq \gamma - p_z = const$

The main contribution to the sum comes from the terms where the phase does not change. The conditions for the stationarity of the phases are the conditions for resonances. Let the term with $n = 0$ be the stationary member. Then the equation for the phase (13) can be replaced by the equation:

$$\dot\Phi = (1-v_z)\left[1 - \frac{\varepsilon}{p_\perp}\sin\Phi\right] \quad (14)$$

Where $\Phi = (\tau - z) + \theta$.

The first bracket on the right-hand side of equation (14) is positive. We consider the relativistic case. In this case, this bracket is small and only decreases with acceleration. If the transverse energy of the particles does not change, then equation (14) resembles the Adler equation known from the theory of synchronization (see, for example, [6,7]).

At $\varepsilon > p_\perp$ there is a stationary state ($\dot\Phi_m = 0$). If $\cos\Phi_m > 0$ this steady state is stable. However, the dynamics of particles is described not only by Eq. (14), but also by equations for transverse and longitudinal momentum. They must be taken into account. So the equation for the longitudinal impulse is:

$$\dot p_z = \varepsilon\frac{p_\perp}{\gamma}\sum_{n=-\infty}^{\infty} J_n(\mu)\cos\left[(\tau - z) + \theta - n\theta\right]$$
(15)

Let us leave only the stationary term in the sum of the right-hand side. We assume that the phase is stationary at $n = 0$. Then equation (15) is simplified:

$$\dot p_z = \varepsilon\frac{p_\perp}{\gamma}J_0(\mu)\cos\Phi_m \quad (16)$$

Taking into account that in our approximation $\mu \sim k_x \ll 1$ we find that the magnitude of the longitudinal impulse depends on time according to the law, resembling the resonant interaction:

$$p_z \approx p_z(0) + (\varepsilon \cdot \cos\Phi_n)\tau \quad (17)$$

The magnitude of the transverse momentum is determined by the equation

$$\dot{p}_\perp = \varepsilon(1-v_z)\sum_{n=-\infty}^{\infty} J_n(\mu)\cos\left[(\tau-z)+\theta-n\theta\right]$$
(18)

The considerations similar to the one used for the defining $p_z$, give:

$$\dot{p}_\perp = \varepsilon(1-v_z)J_0(\mu)\cos\Phi_m \approx$$
$$\approx \left(\frac{1}{2}\varepsilon \cdot k_x \cdot \cos\Phi_m\right)$$
(19)

The magnitude of the transverse momentum also grows linearly with time. However, the slope of this linear increase has a small multiplier:

$$p_\perp \approx p_\perp(0) + \left(\varepsilon\cdot\frac{k_x}{2}\cos\Phi_m\right)\tau \quad (20)$$

When obtaining (20), we took into account what $(1-v_z)\approx(1-k_zv_z-k_xv_x)>0$ and the value of this bracket can be approximated by the value of $k_x$. Note that the numerical calculations are in good agreement with this estimate (see below).

Thus, asymptotically the following time dependences emerge:

$$\gamma \approx p_z \approx \varepsilon\cdot\tau; \quad p_\perp \approx k_x\varepsilon\cdot\tau \quad (21)$$

Let us now return to the equation for phases (14). Taking into account asymptotes (21), this equation can be rewritten:

$$\dot{\Phi} \approx -(1-v_z) \approx \frac{1}{2\gamma^2} \sim \frac{1}{\tau^2} \quad (22)$$

Thus, asymptotically $\Phi_m = const$. The set of results obtained from the analysis of equations (12) - (22) shows that a field of a regular transverse wave in vacuum resonantly accelerates charged particles (electrons) when the conditions, identified above are realized.

3. **Numerical analysis of the original system of equations (5)**

The analytical results obtained above are in many aspects the asymptotic estimates. A series of numerical calculations to obtain the solution of the system of equations (5) was carried out to clarify the conditions under which the resonant acceleration of particles by the field of a transverse electromagnetic wave in vacuum can be achieved. We note from the beginning that good qualitative agreement was obtained between the numerical and analytical results. Typical results of numerical calculations are presented in Fig. 1 - 6.

The main feature of the obtained resonant conditions is that the greater the field strength of the electromagnetic wave $\mathcal{E}$ and the greater the initial longitudinal velocity of the particles $p_z(0)$, the easier the charged particles get into the resonance conditions and the easier they are trapped by the resonance. Fig. 1 - 2 show the results of numerical analysis for the values of the initial conditions and wave parameters that correspond to the onset of particle capture into resonant acceleration. Unlimited acceleration of charged particles is observed. The value of the longitudinal impulse grows linearly during all simulation time (Fig. 1). Moreover, the growth rate of the transverse impulse is well described by the formula (20), and is 10 times lower than the velocity of the longitudinal impulse. This corresponds to the fact that the transverse wavenumber is 10 times smaller than the longitudinal wavenumber. Comparison between the formulas (19) - (21) and the numerical solution shows that there is a good qualitative agreement between them.

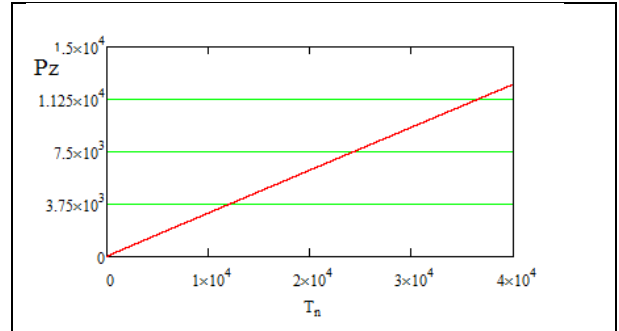

Fig.1. Time dependence of the longitudinal impulse at: $\mathcal{E}$ =2; Pz=2; Px=0.5; $\varphi$ =0.10.

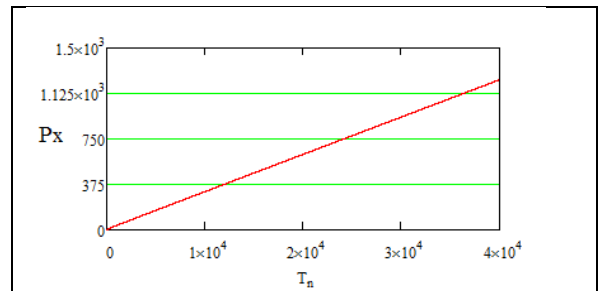

Fig.2. Time dependence of the transverse impulse at: $\mathcal{E}$ =2; Pz=2; Px=0.5; $\varphi$ =0.10.

The saturation process is not present on Fig. 1 - 2. To see the process of transition from unlimited acceleration to a mode in which the acceleration process is limited, it is sufficient to reduce the value of the longitudinal initial impulse to 1.07. This process is shown on Fig. 3. This figure shows that the change of the value of the longitudinal impulse becomes already nonlinear. Some saturation of the particle acceleration process is observed. Thus, if the parameter of the wave force is of the order of 2, then at the value of the initial longitudinal momentum slightly larger than unity, the complete capture of particles into unbounded resonant acceleration does not happen.

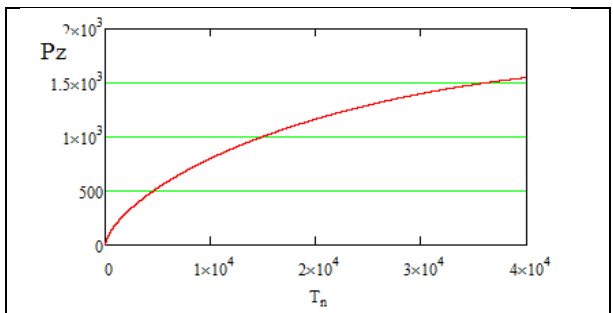

Fig.3. Time dependence of the longitudinal impulse at: $\mathcal{E}$ =2; Pz=1.07; Px=0.0; $\varphi$ =0.10.

The saturation process develops even faster when the value of the transverse wave vector is reduced by another 10 times ($k_x \sim 0.01 \cdot k_z$). This acceleration mode is shown on Fig. 4.

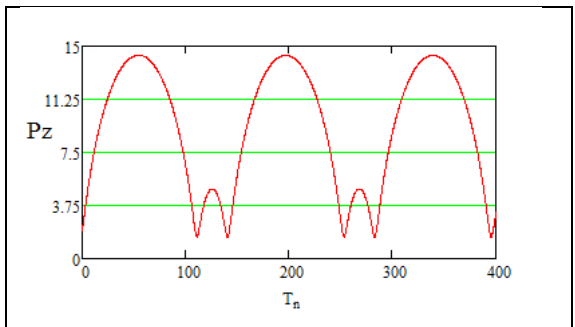

Fig.4. Time dependence of the longitudinal impulse at: $\mathcal{E}$ =2; Pz=2; Px=0.5; $\varphi$ =0.010.

Disruption of the capture of particles in an unlimited resonant acceleration also occur in the case when, at not too high intensities of the external electromagnetic waves, and also at not too high initial values of the longitudinal velocity of the particle, and when its transverse momentum may turn out to be greater than the parameter of the wave force $(p_x > \mathcal{E})$. This situation corresponds to the case when the synchronization process described by the Adler equation does not have stationary stable points. In this case, the synchronization process that promotes the capture of particles into an unlimited resonant acceleration ceases to work, and if the second mechanism of particle capture does not come into play, then the process of resonant unlimited acceleration breaks down (see Fig. 5)..

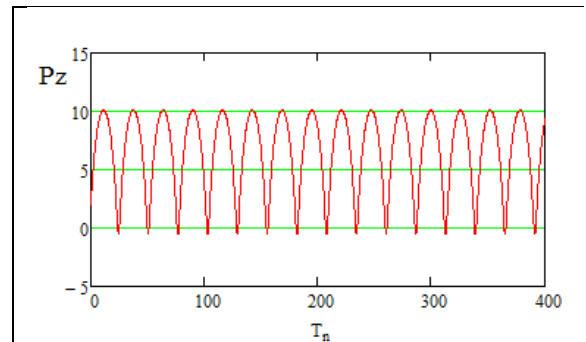

Fig.5. Time dependence of the longitudinal impulse at: $\mathcal{E}$ =2; Pz=2; Px=3; $\varphi$ =0.10.

Increasing the wave force parameter by a factor of two increases the maximum value of the longitudinal impulse by more than two times. This result is shown on Fig. 6. This figure shows the longitudinal momentum as function of time. All parameters of this case equal to the parameters of the case on Fig. 1, except the parameter of the wave strength is doubled ($\mathcal{E} = 4$).

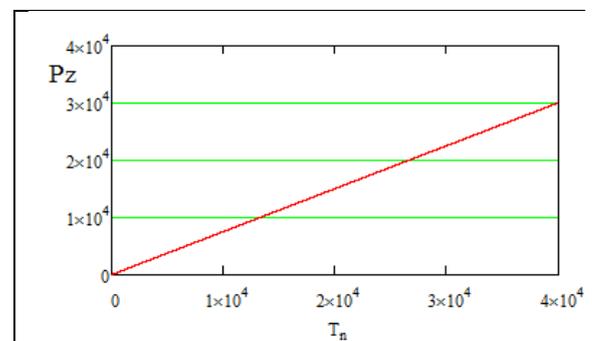

Fig.6. Time dependence of the longitudinal impulse at: $\mathcal{E}$ =4; Pz=2; Px=0.5; $\varphi$ =0.10.

If the transverse wavenumber is not too small, then the analytical results cannot be considered correct. However, there are no limitations for obtaining numerical results. In particular, it turns out that an increase in the angle from 0.1 to 0.6 increases the maximum value of the longitudinal impulse almost six fold. A further increase in the angle leads to a decrease in the value of the longitudinal impulse. This result applies only to particles whose initial values are corresponding to Fig. 1.

One more remark should be made. The capture of a particle by resonant acceleration is easier the greater the parameter of the wave force and the higher the longitudinal momentum of the particle. Moreover, looking at the expression for the longitudinal momentum (the third expression in system (7)), one might think that only when the wave force parameter is greater than unity it is possible to capture particles in the conditions of resonant acceleration. However, in the general case, this is incorrect. In particular, if the longitudinal momentum of the particle is large enough, then capture into resonant acceleration is possible even when the wave force parameter is less than unity. This fact is illustrated on Fig. 7. Note that a decrease in the wave force parameter by five hundredths (to $\mathcal{E}=0.4$) disrupts the capture of particles into resonance.

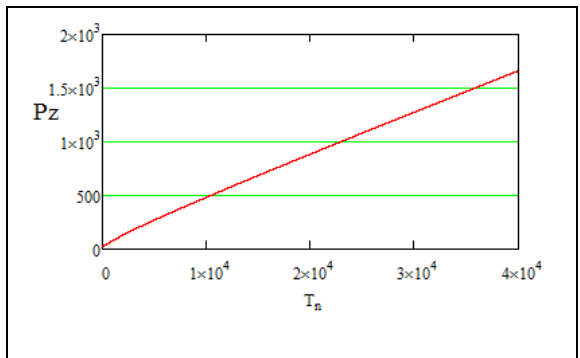

Fig.7. Time dependence of the longitudinal impulse at: $\mathcal{E}$=0.45; Pz=20; Px=0.5; $\varphi$=0.10.

**4. Dynamics of particles in the field of a wave with linear polarization**

The dynamics of particles in the field of a wave with linear polarization described by the system of equations (5) in which it is sufficient to set the strength of the field component $\mathcal{E}_y$ equal to zero ($\mathcal{E}_y=0$). All the features of the dynamics of particles in the field of a wave with linear polarization are qualitatively similar to the features of the dynamics of particles in the field of a wave with circular polarization. For this reason, we do not analyse this dynamics in detail.

**Conclusion**

Let us summarize the most important results obtained in this work:

1. The most important of these results is the result that a transverse electromagnetic wave can resonantly accelerate charged particles in a vacuum. Moreover, the acceleration is performed by both a circularly polarized wave and a linearly polarized wave.

2. The resonant acceleration emerges from the influence of the transverse dynamics of the particles on the phase dynamics of particles themselves. The equations above account for this fact by considering the role of the transverse component of the wave vector $k_x$ on the dynamics of the particle. Note the difference (in our cases) between the fact that the wave moves at an angle to the z-axis and the fact that the particle can move at a certain angle with respect to the wave vector of the wave. These two cases are fundamentally different. In the second case, when the particle moves at an angle to the wave vector of the wave, the phase characteristics do not depend on the transverse motion of the particles. In this case, particles are not captured into a resonance. It should be noted that a similar situation occurs when cyclotron resonances arise. If the wave propagates strictly along the z axis, along the direction of the external magnetic field, then only the autoresonance conditions are realized. Ordinary cyclotron resonances appear only when the wave has a nonzero transverse component of the wave vector (see, for example, [8,9]).

3. The condition for the capture of charged particles in the resonant acceleration mode is the presence of a wave with the large enough force parameter as well as the presence of a sufficiently large initial longitudinal momentum of the charged particle ($p_z > 1$). The larger these parameters are, the easier is to capture particles into the resonant acceleration mode.

4. Note two features that characterize the capture of particles in the resonant acceleration mode. The first feature is associated with the fact that the phase dynamics of particles at the initial stage of the acceleration process is described by an equation that resembles the Adler equation in the theory of synchronization. However, Adler's equation contains functions that are independent of time. Equation (14), which resembles Adler's equation, contains functions that change over time. For this reason, the synchronization process can only be carried out for a limited time interval. The second feature of the capture of charged particles in the resonant acceleration mode is associated with the first factor on the right-hand side of Eq. (14). In the relativistic case this factor is small and positive. In addition, it decreases rather quickly. It is easy to see that the factor is inversely proportional to the square of the particle's energy. Considering that the energy grows linearly with time, the phase derivative quickly tends to zero (as $1/\tau^2$). The phase itself tends to a constant value rather quickly. Moreover, an analysis of the numerical results shows that the stationary phase itself tends to zero.

ACKNOWLEDGMENTS
This work was supported by the Target Program "Physics of Plasmas and Plasma Electronics: fundamentals and applications" of the National Academy of Sciences of Ukraine (Grant No. 0117U006867)

# О ВОЗМОЖНОСТИ РЕЗОНАНСНОГО УСКОРЕНИЯ ЗАРЯЖЕННЫХ ЧАСТИЦ ПОЛЕМ ПОПЕРЕЧНОЙ ЭЛЕКТРОМАГНИТНОЙ ВОЛНЫ В ВАКУУМЕ

Изложены результаты аналитического и численного исследования динамики заряженных частиц в поле поперечной электромагнитной волны большой амплитуды в вакууме. Показано, что существуют условия, при выполнении которых происходит резонансное ускорение частиц. Эти условия сформулированы. Обсуждаются особенности и механизм этого ускорения
PACS: 41.75.Jv; 52.38.Kd; 96.50.Pw

1. **Введение**

Ускорение заряженных частиц интенсивными электромагнитными полями в вакууме является заманчивой целью многих исследований. Достоинства такого ускорения хорошо известны. Известны также и трудности, с которыми сталкиваются исследователи при попытках достичь этой цели. Достаточно сослаться на работы [1-3] и на цитируемую в них литературу, чтобы уяснить всю сложность проблемы, ее привлекательность, а также состояние дел в этом направлении исследований. Работа состоит из четырех разделов и заключения. Во втором разделе описана динамика частиц в поле волны с круговой поляризацией. В третьем разделе описаны результаты численного анализа этой динамики. В четвертом разделе речь идёт о динамике частиц в поле с линейной поляризацией. В заключении сформулированы основные результаты и дано обсуждение некоторых из них.

**2. Постановка задачи и основные уравнения**

в поле плоской электромагнитной волны, которая в общем случае имеет следующие компоненты:
$$\mathcal{E} = \mathrm{Re}(\mathbf{E}\exp(i\omega t - i\mathbf{kr})),$$
$$\mathbf{H} = \mathrm{Re}\left(\frac{c}{\omega}[\mathbf{kE}]\exp(i\omega t - i\mathbf{kr})\right) \quad (1)$$

где $\mathbf{E} = \mathbf{a}\cdot E$, $\mathbf{a} = \{\alpha_x, i\alpha_y, \alpha_z,\}$ — вектор поляризации волны.
Без ограничения общности можно считать, что волновой вектор $\mathbf{k}$ имеет только две отличные от нуля компоненты $k_x$ и $k_z$. В безразмерных переменных $\mathbf{p} \to \mathbf{p}/mc$, $\tau \to \omega t$, $\mathbf{r} \to \mathbf{r}(\omega/c)$ уравнения движения частицы можно привести к виду.

$$\frac{d\mathbf{p}}{d\tau} = \left(1 - \frac{\mathbf{kp}}{\gamma}\right)\mathrm{Re}\left(\mathcal{E}e^{i\psi}\right) + \frac{\mathbf{k}}{\gamma}\mathrm{Re}\left[(\mathbf{p}\mathcal{E})e^{i\psi}\right] \quad (2)$$

$$\mathbf{v} = \frac{d\mathbf{r}}{d\tau} = \frac{\mathbf{p}}{\gamma}, \quad \dot{\psi} = \frac{d\psi}{d\tau} = 1 - \frac{\mathbf{kp}}{\gamma},$$

Где $\mathcal{E} = e\mathbf{E}/mc\omega$; $\psi = \tau - \mathbf{kr}$, $\mathbf{k}$ — единичный вектор в направлении распространения волны, $\gamma = \sqrt{1+\mathbf{p}^2}$ — энергия частицы, $\mathbf{p}$ – ее импульс. Помножив первое из уравнения (2) на $\mathbf{p}$, получим следующее уравнение, описывающие изменение энергии частицы:

$$\frac{d\gamma}{d\tau} = \mathrm{Re}\left(\mathbf{v}\mathcal{E}e^{i\psi}\right) \quad (3)$$

Используя уравнения (3), из системы уравнения (2) находим интеграл движения

$$\mathbf{p} - \mathrm{Re}\left(i\mathcal{E}e^{i\psi}\right) - \mathbf{k}\gamma = \mathbf{const} \equiv \mathbf{C}. \quad (4)$$

2. **Динамика частиц в поле волны с круговой поляризацией**

Динамика частиц в поле волны с круговой поляризацией наиболее полно отображает все особенности интересующей нас динамики. Поэтому мы, прежде всего, рассмотрим именно эту динамику. Будем считать, что волновой вектор электромагнитной волны расположен в плоскости $\{x,z\}$, то есть $\mathbf{k} = \{k_x, 0, k_z\}$. Тогда система уравнений (2) для описания

движения частиц в поле волны с круговой поляризацией приобретает вид:

$$\dot{p}_x = \mathcal{E}_x \left(1 - k_x v_x - k_z v_z\right) \cos\psi +$$
$$+ \frac{k_x}{\gamma}\left[\left(\mathcal{E}_x p_x + \mathcal{E}_z p_z\right)\cos\psi - \mathcal{E}_y p_y \sin\psi\right]$$
$$\dot{p}_y = -\mathcal{E}_y \left(1 - k_x v_x - k_z v_z\right)\sin\psi$$
$$\dot{p}_z = \mathcal{E}_z \left(1 - k_x v_x - k_z v_z\right)\cos\psi +$$
$$+ \frac{k_z}{\gamma}\left[\left(\mathcal{E}_x p_x + E_z p_z\right)\cos\psi - \mathcal{E}_y p_y \sin\psi\right]$$
$$\dot{\gamma} = \frac{1}{\gamma}\left[\left(\mathcal{E}_x p_x + \mathcal{E}_z p_z\right)\cos\psi - \mathcal{E}_y p_y \sin\psi\right] \quad (5)$$

Здесь $\dot{p} \equiv \dfrac{dp}{d\tau}$

В действительности система уравнений (5) является системой уравнений, которая описывает динамику частиц в поле волны с эллиптической поляризацией. Однако нас будет интересовать динамика частиц в поле волны с круговой поляризацией. Поэтому ниже мы будем считать, что $\mathcal{E}_z = 0,\ \mathcal{E}_x = \mathcal{E}_y = \mathcal{E}$.

Рассмотрим вначале случай, когда волна с круговой поляризацией распространяется строго вдоль оси $z$
($k_x = \sin\varphi = 0,\ k_z = \cos\varphi = 1$).

$$\dot{p}_x = \dot{\psi}\cdot\mathcal{E}\cdot\cos\psi$$
$$\dot{p}_y = -\dot{\psi}\cdot\mathcal{E}\cdot\sin\psi \quad (6)$$

Решением системы уравнений (6) будут функции:

$$p_x = \mathcal{E}\sin\psi + F_0 = \gamma v_x = \frac{C_z}{\dot{\psi}}\dot{x} = C_z \frac{dx}{d\psi}$$

$$p_y = \mathcal{E}\cos\psi + \Phi_0 = \gamma v_y = \frac{C_z}{\dot{\psi}}\dot{y} = C_z \frac{dy}{d\psi} \quad (7)$$

$$p_z = p_z(0) + \frac{E^2}{|C_z|}\left[1 - \cos(\psi - \psi_0)\right]$$

Где $F_0 = p_x(0) - \mathcal{E}\sin\psi_0$;
$\Phi_0 = p_y(0) - \mathcal{E}\cos\psi_0$, $p_z - \gamma = C_z = const$ -- интеграл. Нулевые индексы соответствуют начальным значениям функций.

Отметим, что аналогичные решения были получены еще в работе [4]. Обратим внимание на выражения (7). Из них следует, что поперечная динамика частиц пропорциональна первой степени параметра силы волны, а продольная динамика – пропорциональна квадрату этого параметра. Таким образом, параметр $\mathcal{E} = 1$ разделяет качественно разную динамику частиц. При $\mathcal{E} < 1$ преобладает поперечная (привычная) динамика частиц. Если же $\mathcal{E} > 1$, то преобладает продольная динамика частиц. Обратим также внимание, что второе слагаемое в выражении для продольного импульса обратно пропорционально интегралу ($\gamma - p_z = -C_z = const$). Во многих случаях этот интеграл может быть малым. При этом продольная динамика частиц приобретает дополнительное преимущество по отношению к поперечной динамике.

Рассмотрим структуру поперечного движения частиц. Для этого выпишем выражения для поперечных координат частиц:

$$(x - x_0) = -\frac{\mathcal{E}}{C_z}\cos\psi + F_0\cdot\psi + \frac{1}{C_z}\mathcal{E}\cdot\cos\psi_0$$
$$(y - y_0) = \frac{\mathcal{E}}{C_z}\sin\psi + \Phi_0\cdot\psi - \frac{1}{C_z}\mathcal{E}\cdot\sin\psi_0 \quad (8)$$

Для определенности выберем следующие значения начальных функций:

$$\psi_0 = 0,\ F_0 = \Phi_0 = 0,\ p_x(0) = 0,\ p_y(0) = \mathcal{E} \quad (9)$$

Тогда частицы в поле такой волны будут описывать окружности в своем поперечном сечении:

$$(x - x_1)^2 + (y - y_0)^2 = \left(\frac{\mathcal{E}}{C_z}\right)^2 \quad (10)$$

Таким образом, траектория частиц в поперечном сечении может представлять окружность. Наиболее интересным для нас является тот факт, что поперечная динамика частиц не влияет на фазовую динамику частиц. Такое влияние возникает только в том случае, когда волна распространяется под некоторым углом к оси Z, когда поперечное волновое число отлично от нуля ($k_x = \sin\varphi \neq 0$).

Если у волны с круговой поляризацией поперечная компонента волнового вектора отлична от нуля, то такие простые выражения, которые описывают динамику частиц, получить не удается. Однако, если эта компонента волнового вектора мала ($k_x \ll 1$), то удается обнаружить новые важные особенности динамики частиц. В этом случае, удобно ввести новые переменные:

$$p_x = p_\perp \cos\theta, \quad p_y = p_\perp \sin\theta, \quad p_z = p_\parallel,$$
$$p_\perp = \sqrt{p_x^2 + p_y^2}, \quad x = \xi - \frac{p_\perp}{\gamma\dot\psi}\sin\theta, \quad (11)$$
$$y = \eta + \frac{p_\perp}{\gamma\dot\psi}\cos\theta$$

В этих новых переменных явно учтена осцилляторная динамика частиц в поперечном направлении. Поперечная динамика и фазовая динамика в этих новых переменных описывается уравнениями

$$\dot p_\perp = \dot p_x \cos\theta + \dot p_y \sin\theta$$
$$\dot\theta = \frac{1}{p_\perp}\left(\dot p_y \cos\theta - \dot p_x \sin\theta\right) \quad (12)$$

$$\psi = (\tau - k_z z - k_x \xi) - \mu \sin\theta \equiv a - \mu \sin\theta,$$
где $\mu = k_x p_\perp / \gamma \cdot \dot\psi$

Для дальнейшего удобно воспользоваться формулами разложения (смотри, например, [5])

$$\cos\psi = \cos(a - \mu\sin\theta) = \sum_{n=-\infty}^{\infty} J_n(\mu)\cos(a - n\theta)$$

$$\sin\psi = \sin(a - \mu\sin\theta) = \sum_{n=-\infty}^{\infty} J_n(\mu)\sin(a - n\theta)$$

Учтем, что $k_x \ll 1; (k_z - 1) \ll 1$. При этом оказывается, что новые результаты можно получить учитывая величину $k_x$ только в выражениях для фаз. Тогда второе уравнение (уравнение для фаз) системы (12) приобретает вид:

$$\dot\theta = -\varepsilon \frac{(1 - v_z)}{p_\perp} \sum_{n=-\infty}^{\infty} J_n(\mu) \sin[(\tau - z) + \theta - n\theta]$$
(13)

Здесь $\mu \ll 1$, $\gamma\dot\psi \simeq \gamma - p_z = const$

Основную роль в сумме будут играть те члены, у которых фаза меняться не будет. Условия стационарности фаз и будут условиями резонансов. Пусть стационарным членом будет член с $n = 0$. Тогда уравнение для фазы (13) можно заменить уравнением:

$$\dot\Phi = (1 - v_z)\left[1 - \frac{\varepsilon}{p_\perp}\sin\Phi\right] \quad (14)$$

Где $\Phi = (\tau - z) + \theta$

Первая скобка в правой части уравнения (14) положительна. Будем рассматривать релятивистский случай. В этом случае она мала и при ускорении только уменьшается. Если поперечная энергия частиц не меняется, то уравнение (14) напоминает уравнение Адлера в теории синхронизации (смотри, например, ). При $\varepsilon > p_\perp$ имеется стационарной состояние ($\dot\Phi_m = 0$). Если $\cos\Phi_m > 0$ это стационарное состояние будет устойчивым. Однако динамика частиц описывается не только уравнением (14), но и уравнениями для поперечного и продольного импульса. Их надо учитывать.

Так уравнение для продольного импульса имеет вид:

$$\dot p_z = \varepsilon \frac{p_\perp}{\gamma} \sum_{n=-\infty}^{\infty} J_n(\mu) \cos[(\tau - z) + \theta - n\theta] \quad (15)$$

Оставим в сумме правой части только стационарный член. Будем считать, что стационарность фазы наступает при $n = 0$. Тогда уравнение (15) упрощается:

$$\dot p_z = \varepsilon \frac{p_\perp}{\gamma} J_0(\mu) \cos\Phi_m \quad (16)$$

Учитывая, что в нашем приближении $\mu \sim k_x \ll 1$ найдем, что величина продольного импульса зависит от времени по закону, который характерен для резонансов:

$$p_z \approx p_z(0) + (\mathcal{E} \cdot \cos\Phi_n)\tau \quad (17)$$

Величина поперечного импульса определяется уравнением

$$\dot{p}_\perp = \mathcal{E}(1-v_z)\sum_{n=-\infty}^{\infty} J_n(\mu)\cos[(\tau-z)+\theta-n\theta] \quad (18)$$

Аналогичные соображения, которые были использованы для определения $p_z$, дают:

$$\dot{p}_\perp = \mathcal{E}(1-v_z)J_0(\mu)\cos\Phi_m \approx$$
$$\approx \left(\frac{1}{2}\mathcal{E} \cdot k_x \cdot \cos\Phi_m\right) \quad (19)$$

Величина поперечного импульса также растет со временем по линейному закону. Однако наклон этой линейной функции имеет малый множитель $k_x \ll 1$:

$$p_\perp \approx p_\perp(0) + \left(\mathcal{E} \cdot \frac{k_x}{2}\cos\Phi_m\right)\tau \quad (20)$$

При получении (20) мы учли, что $(1-v_z) \approx (1-k_z v_z - k_x v_x) > 0$ и что можно оценить величину этой скобки величиной $k_x$. Отметим, что численные расчеты хорошо согласуются с этой оценкой (смотри ниже).

Таким образом, асимптотически имеются такие зависимости от времени

$$\gamma \approx p_z \approx \mathcal{E}\cdot\tau; \quad p_\perp \approx k_x \mathcal{E}\cdot\tau \quad (21)$$

Вернемся теперь к фазовому уравнению (14). С учетом асимптотик (21) это уравнение можно переписать:

$$\dot{\Phi} \approx -(1-v_z) \approx \frac{1}{2\gamma^2} \sim \frac{1}{\tau^2} \quad (22)$$

Таким образом асимптотически $\Phi_m = const$. Совокупность результатов, полученных из анализа уравнений (12) – (22) указывают на то, что в рамках сформулированных условий реализуется резонансное ускорение заряженных частиц (электронов) полем регулярной поперечной волны в вакууме.

## 3. Численный анализ исходной системы уравнений (5)

Полученные выше аналитические результаты во многом носят оценочный, асимптотический характер. Для уточнения условий, при которых реализуется резонансное ускорение частиц полем поперечной электромагнитной волны в вакууме, была проведена серия численных расчетов системы уравнений (5). Отметим сразу, что было получено хорошее качественное согласие численных и аналитических результатов. Характерные результаты численных расчетов представлены на рисунках 1 - 6.

. Основная особенность полученных резонансных условий заключается в том, что чем больше напряженность поля электромагнитной волны $\mathcal{E}$ и чем больше начальная продольная скорость частиц $p_z(0)$, тем легче заряженные частицы попадают в условия резонанса. Тем легче они захватываются в резонанс.

На рисунках 1 - 2 приведены результаты численного анализа при значениях начальных условий и параметров волны, которые соответствуют началу захвата частиц в резонансное ускорение. Видно неограниченное ускорение заряженных частиц. Величина продольного импульса растет по линейному закону в течение всего времени счёта (рисунок 1). Причём скорость роста поперечного импульса $p_x$ находится в соответствии с формулой (20), то есть она в 10 раз меньше чем скорость продольного импульса. Это соответствует тому факту, что поперечное волновое число $k_x$ в 10 раз меньше величины продольного волновая числа $k_x \sim 0.1 \cdot k_z$. Сравнивая результаты формул (19) – (21) с результатами численного счёта можно утверждать, что имеется хорошее качественное согласие между ними.

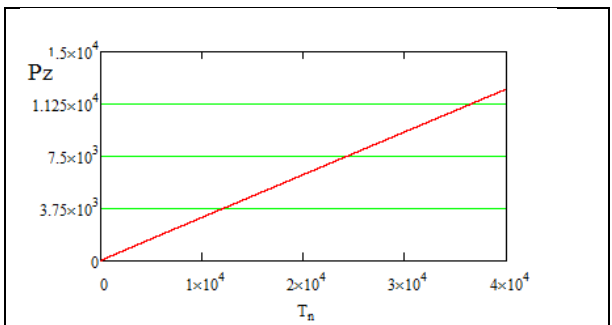

Fig.1. Зависимость продольного импульса от времени при: $\varepsilon=2$; Pz=2; Px=0.5; $\varphi=0.10$

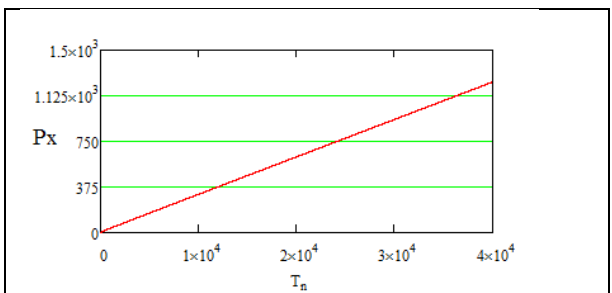

Fig.2. Зависимость поперечного импульса от времени при: $\varepsilon=2$; Pz=2; Px=0.5; $\varphi=0.10$

На рисунках 1 - 2 не виден процесс насыщения. Чтобы увидеть процесс перехода от неограниченного ускорения к режиму, в котором происходит ограничение процесса ускорения достаточно уменьшить величину продольного начального импульса до величины 1.07 . Такой процесс представлен на рисунке 3. На этом рисунке видно, что закон изменения величины продольного импульса становится уже нелинейным. Наблюдается некоторое насыщение процесса ускорения частицы. Таким образом, если параметр силы волны будет порядка 2, то при значении начального продольного импульса слегка большего единицы полного захвата частиц в неограниченное резонансное ускорение не происходит.

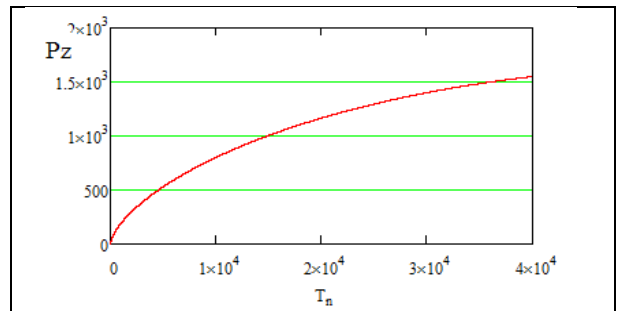

Fig.3. Зависимость продольного импульса от времени при: $\varepsilon=2$; Pz=1.07; Px=0.0; $\varphi=0.10$ Suturation

Процесс насыщения развивается ещё быстрее в том случае, когда величина поперечного волнового вектора уменьшить ещё в 10 раз ( $k_x \sim 0.01 \cdot k_z$ ). Такой режим ускорения представлен на рисунке 4.

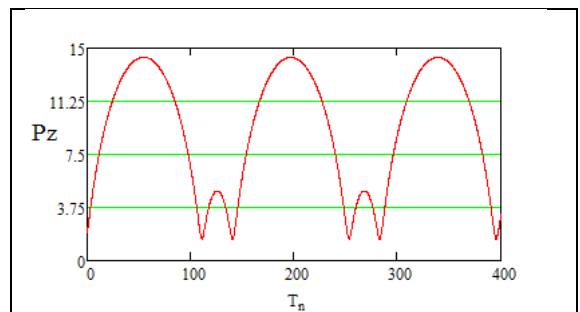

Fig.4. Зависимость продольного импульса от времени при: $\varepsilon=2$; Pz=2; Px=0.5; $\varphi=0.010$

Срыв захвата частиц в неограниченное резонансное ускорение будет происходить и в том случае, когда при не слишком больших напряженностях внешней электромагнитные волны, а также не при слишком больших начальных значениях продольной скорости частицы, её поперечный импульс может оказаться больше чем параметр силы волны $(p_x > \varepsilon)$. Это ситуация соответствует тому случаю, когда процесс синхронизации, который описывается уравнением Адлера не имеет стационарных устойчивых точек. В этом случае процесс синхронизации, который способствует захвату частиц в неограниченное резонансное ускорение перестаёт работать и если в игру не вступает второй механизм захвата частиц, то процесс резонансного неограниченного ускорения срывается (смотри рисунок 5).

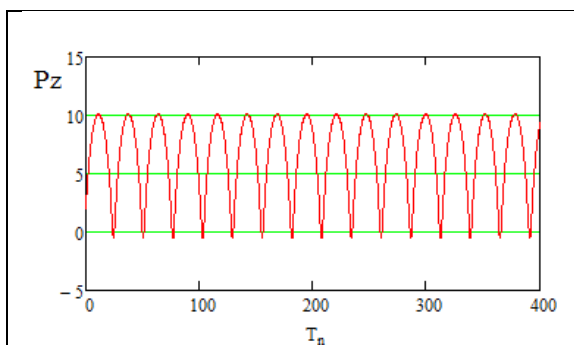

Fig.5. Зависимость продольного импульса от времени при: $\varepsilon$=2; Pz=2; Px=3; $\varphi$=0.10

Увеличение параметра силы волны в два раза увеличивает максимальную величину продольного импульса более чем два раза. Этот результат демонстрируется фигурой 6. На этой фигуре представлена зависимость продольного импульса от времени. Все параметры этого случая совпадают с параметрами случая, который представлен на рисунке 1, за исключением параметра силы волны, который был увеличен в два раза ($\varepsilon = 4$).

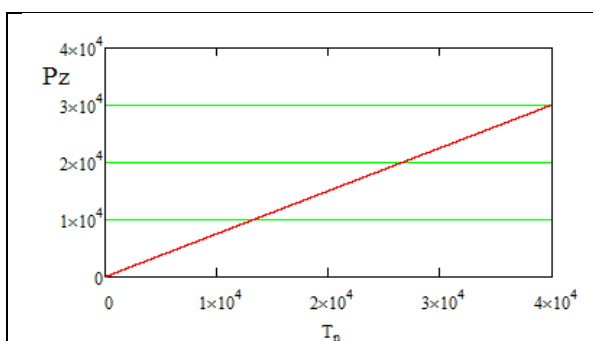

Fig.6. Зависимость продольного импульса от времени при: $\varepsilon$=4; Pz=2; Px=0.5; $\varphi$=0.10

Если поперечное волновое число оказывается не слишком малым та аналитические результаты нельзя считать корректными. Однако для получения численных результатов ограничений нет. В частности, оказывается, что увеличение угла $\varphi$ от величины 0.1 до величины 0.6 увеличивает максимальное значение продольного импульса практически в шесть раз. Дальнейшее увеличение угла $\varphi$ ведёт к уменьшению величины продольного импульса. Этот результат относится только к частицам, начальные значения которых соответствуют рисунку 1.

Следует сделать еще одно замечание. Захват частицы в условия резонансного ускорения тем легче осуществить, чем больше будет параметр силы волны и чем больше будет продольный импульс частицы. Более того, глядя на выражение для продольного импульса (третье выражение в системе (7)) можно подумать, что только при параметре силы волны большем единицы возможен захват частиц в условиях резонансного ускорения. Однако в общем случае это не так. В частности, если продольный импульс частицы достаточно велик, то захват в резонансное ускорение возможен даже при параметре сила волны меньшем единицы. Этот факт иллюстрируется рисунком 7. Отметим, что уменьшение параметра силы волны на пять сотых ($\varepsilon = 0.4$) срывает захват частиц в резонанс

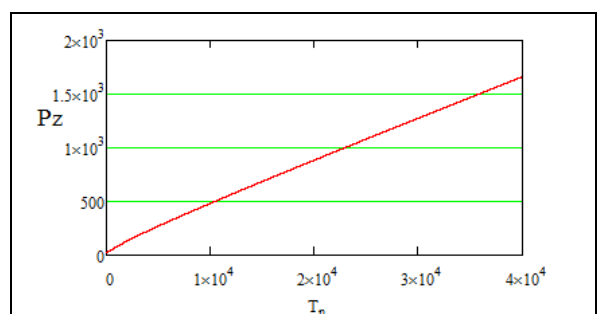

Fig.7. Зависимость продольного импульса от времени при: $\varepsilon$=0.45; Pz=20; Px=0.5; $\varphi$=0.10

## 4. Динамика частиц в поле волны с линейной поляризацией

Динамика частиц в поле волны с линейной поляризацией будет отписываться системой уравнений (5) в которой достаточно положить напряженность компоненты поля $\varepsilon_y$ равной нулю ($\varepsilon_y = 0$). Все особенности динамики частиц в поле волны с линейной поляризацией качественно аналогичны особенностям динамики частиц в поле волны с круговой поляризацией. По этой причине мы не будем детально останавливаться на этой динамике.

Заключение

Отметим наиболее важные результаты, полученные в этой работе:

1. Наиболее важным из этих результатов является результат о том, что в вакууме поперечная электромагнитная волна может резонансно ускорять заряженные частицы. Причём ускорение производится как волной с круговой поляризацией, так и волной с линейной поляризацией.
2. Механизм резонансного ускорения определяется тем фактом, что в рассмотрении учтено влияние на фазовую динамику частиц поперечной динамики самих частиц. Выше этот факт выражался в том, что мы учитывали роль поперечной компоненты волнового вектора $k_x$ на динамику частицы. Следует отметить разницу между тем, что волна движется под углом к оси z и тем, что частица может двигаться под некоторым углом по отношению к волновому вектору волны. Эти два случая принципиально различаются. Во втором случае, когда частица движется под углом к волновому вектору волны, фазовые характеристики не зависят от поперечного движения частиц. Захват частиц в резонанс в этом случае не происходит. Следует заметить что аналогичная ситуация имеет место и при возникновении циклотронных резонансов. Если волна распространяется строго вдоль оси z, вдоль направления внешнего магнитного поля, то реализуются только условия авторезонанса. Обычные циклотронные резонансы появляется только в том случае, когда у волны имеется отличный от нуля поперечный компонент волнового вектора (смотри, например, [8,9] )
3. Условием захвата заряженных частиц в режим резонансного ускорения является наличие волны , параметр силы которой достаточно большой, а также наличие достаточно большого начального продольного импульса заряженной частицы ($p_z > 1$). Чем большим будут эти параметры, тем легче происходит захват частиц в режим резонансного ускорения.
4. Отметим две особенности, которые характеризуют захват частиц в режим резонансного ускорения. Первая особенность связано с тем что фазовая динамика частиц на начальном этапе процесса ускорения описывается уравнением, которое напоминает уравнение Адлера в теории синхронизации. Однако уравнение Адлера содержит функции, которые не зависят от времени. Уравнение (14), которое напоминает уравнение Адлера, содержит функции, которые меняются во времени. По этой причине процесс синхронизации может осуществляться только на ограниченном интервале времени. Вторая особенность захвата заряженных частиц в режим резонансного ускорения связано с первым множителем правой частью уравнения (14). В релятивистском случае этот множитель мал. Он положителен. Кроме того, он достаточно быстро уменьшается. Легко увидеть, что он обратно пропорционален квадрату энергии частицы. Учитывая, что энергия растёт с ростом времени по линейному закону, то производная фазы быстро стремится к нулю (как $1/\tau^2$). Сама фаза достаточно быстро стремится к постоянной величине. Более того, анализ численных результатов показывает, что стационарная фаза сама стремиться к нулю.

ACKNOWLEDGMENTS

This work was supported by the Target Program "Physics of Plasmas and Plasma Electronics: fundamentals and applications" of the National Academy of Sciences of Ukraine (Grant No. 0117U006867)